\documentclass[aps,prb,preprint,12pt]{revtex4}
\usepackage{bm}

\begin{document}

\title{Vector constants of the motion and orbits in the Coulomb/Kepler
problem}
\author{Gerardo Mu\~{n}oz}
\affiliation{Department of Physics,
California State University Fresno,
Fresno, California 93740-0037}

%\date{\today}

\begin{abstract}
The equation for the conic sections describing the possible orbits in 
a potential $V \sim r^{-1}$ is
obtained by means of a vector constant of the motion differing from 
the traditional Laplace-Runge-Lenz vector.
\end{abstract}

\maketitle

The Laplace-Runge-Lenz\cite{Laplace,Runge,Lenz,Goldstein1} (or 
Hermann-Bernoulli-Laplace\cite{Goldstein2}) vector provides a
simple and elegant way of obtaining the equation for the orbit of a 
particle moving under the influence of a radial inverse-square-law
force.\cite{GorringeLeach} This method also has the added advantage over the more
common approach of direct integration  of the  equations
of motion\cite{MarionThornton} that the very existence of this 
single-valued constant of the motion explains the
otherwise surprising degeneracy of the Coulomb/Kepler 
problem.\cite{Fradkin} Our aim here is to show that there is another
vector  constant of the motion which makes a first principles derivation
of the equation  of the orbit even easier than the Laplace-Runge-Lenz
vector or the direct integration methods.

The force law ${\bf F} = -k {\bf \hat r}/r^2$ (we assume $k>0$; the 
repulsive case may be treated along the same lines) leads to the
nonrelativistic equation of motion
\begin{equation}
{d {\bf p} \over dt}  =  -k {{\bf \hat r} \over r^2},
\end{equation}
and to the well-known conservation laws for energy $E = \frac12 m v^2 
- k/r$ and angular momentum ${\bf L} = {\bf r} \times {\bf p}$.
If we choose polar coordinates in the plane of the orbit, we have the 
usual expression $L = mr^2 \dot\theta$, and also
\begin{equation}
\label{2}
{d {\bf p} \over dt}  =  m {d {\bf v} \over 
d\theta}\dot\theta  = {d {\bf v} \over d\theta} {L \over r^2} 
\, .
\end{equation}
Combining  Eq.~(\ref{2}) with Eq.~(1) we obtain
\begin{equation}
{d {\bf v} \over d\theta}  =  - {k \over L} {\bf \hat r} \,.
\end{equation}
Note that Eq.~(3) implies immediately that ${\bf v}(0) = {\bf 
v}(2\pi)$ and that the velocity vector traces out a circle if $\theta$
is allowed to vary from $0$ to $2\pi$. Integration of Eq.~(3) is 
trivial if we recall that ${d {\bm {\hat\theta}}/ d\theta} = -{\bf 
\hat r}$:
\begin{equation}
{\bf v}  -  {k \over L} {\bm {\hat\theta}}  = {\bf u} \,,
\end{equation}
where ${\bf u}$ is a constant vector. This vector is the 
constant of the motion we propose to  use instead of the
Laplace-Runge-Lenz vector in the derivation of the orbits. It is
interesting to note that a quaternion equivalent to  Eq.~(4) was already
known to Hamilton in 1845.\cite{Hamilton}  Unfortunately,
this simple result seems to have vanished from textbooks on classical 
mechanics after the first decade of the twentieth
century.\cite{AbelsondiSessaRudolph}

The vector ${\bf u}$ may be evaluated in terms of the physical 
parameters of the problem. If we choose the angle so that the minimum value of $r$ is at $\theta = 0$ and adjust
our coordinate system so that ${\bf v}  =  v_0 {\bf \hat y}$ at $r_{\rm min}$, we have
\begin{equation}
{\bf u}   =  (v_0 - {k \over L}) {\bf \hat y} \,.
\end{equation}
Because $L = mr_{\rm min}v_0$, the energy is $E = \frac12 m v_0^2 - 
k/r_{\rm min} = \frac12 m v_0^2 - kmv_0/L$. We solve this equation for
$v_0$ and obtain
\begin{equation}
v_0  =  {k \over L} \pm \sqrt{\left({k \over L}\right)^2 + {2E 
\over m}}  = {k \over L}(1\pm\epsilon)\,,
\end{equation}
where $\epsilon = \sqrt{1 + {2EL^2 / mk^2}}$ will turn out to be the 
eccentricity. The plus sign is required at $r_{\rm min}$; substituting this
result into Eq. (5) provides  us with an
alternative form for ${\bf u}$:
\begin{equation}
{\bf u}   =  {k \over L} \epsilon {\bf \hat y} \,.
\end{equation}

Obtaining the equation of the orbit is now straightforward. By taking 
the scalar product of Eq.~(4) with ${\bm {\hat\theta}}$ and using
${\bf v} \cdot {\bm {\hat\theta}} = r \dot \theta = L/mr$, ${\bf \hat 
y} \cdot {\bm {\hat\theta}} = \cos \theta$, we find
\begin{equation}
{L \over mr}  -  {k \over L}  =  {k \over L} \epsilon 
\cos \theta \,.
\end{equation}
The definition $\alpha = L^2/mk$ allows us to write the solution for $r$ as
\begin{equation}
r  =  {\alpha \over {1+\epsilon \cos \theta}} \,,
\end{equation}
which is the usual equation of a conic section with one focus at the
origin  and eccentricity $\epsilon$. Note that the velocity
${\bf v}   =  k [\epsilon \sin \theta {\bf \hat r}  +  
(1+\epsilon \cos \theta){\bm {\hat\theta}}]/L
 =  k [- \sin \theta {\bf \hat x}  +  (\epsilon + \cos 
\theta){\bm {\hat y}}]/L$ follows almost
trivially from Eqs.~(4) and (7). Hence the orbit in {\it
velocity space} is always circular and characterized by
$v_x^2 +(v_y -k \epsilon /L)^2 = k^2/L^2$. More precisely, the orbit 
in velocity space is a  circle of radius $k/L$ and center at
$(0, k\epsilon/L)$ if the orbit
in position space is a circle ($\epsilon = 0$), an ellipse 
($0<\epsilon<1$), or a parabola ($\epsilon = 1$). On the other hand, 
if the spatial
trajectory is a hyperbola ($\epsilon>1$), the angle ranges from 
$-\cos^{-1}(-1/\epsilon)$ to $\cos^{-1}(-1/\epsilon)$ only, and the 
velocity space
orbit is a section of a circle in the upper $v_x, v_y$ plane with 
$k(\epsilon^2-1)/\epsilon L < v_y \leq k(\epsilon+1)/L$.

The vector  constant of the motion ${\bf u}$ is simpler (and its
derivation certainly  easier\cite{Kaplan}) than the Laplace-Runge-Lenz
vector, but the two  are,
of course, not independent constants of
 the motion. Indeed, the cross product of
Eq.~(4) with $m {\bf L}$ yields
\begin{equation}
{\bf p} \times {\bf L}  -  mk {\bf \hat r} =  m{\bf u} 
\times {\bf L}  .
\end{equation}
The left-hand side is the Laplace-Runge-Lenz vector ${\bf A}$, so 
${\bf A} =  m{\bf u} \times {\bf L}$. The three conserved vectors 
${\bf A}$,
${\bf u}$, and ${\bf L}$ form a right-handed orthogonal 
system equivalent to ${\bf \hat x}$, ${\bf \hat y}$, and ${\bf \hat 
z}$ after
normalization.

As a final comment of pedagogical interest, we present a second 
approach that may be useful in an elementary discussion of the
Coulomb/Kepler problem. We begin with the expression for the energy 
$E = \frac12 m v^2 - k/r$.
If we use ${\bf v}\cdot{\bm{\hat\theta}} = r\dot\theta = L/mr$ to
replace 
$1/r$ by $m{\bf v}\cdot{\bm{\hat\theta}}/L$ in the potential
term, we may write $E$ as
\begin{equation}
E  =  \frac12 m v^2 - {mk \over L} {\bf v}\cdot {\bm {\hat\theta}}
\, .
\end{equation}
If we complete the square in the velocity, we find
\begin{equation}
E  =  \frac12 m u^2 - {mk^2 \over 2L^2}\,,
\end{equation}
with ${\bf u}$ the same vector as in Eq.~(4). It follows that  the
magnitude of ${\bf u}$ must be constant. To prove that this  vector is
conserved both in magnitude and direction, we must only take a 
derivative of ${\bf u}$ and substitute $d{\bf v}/dt = -k {\bf \hat 
r}/mr^2$ and
$d {\bm {\hat\theta}}/dt = -\dot \theta{\bf \hat r}$ into the result. 
The equation of the orbit may then be found as before. This approach 
does not require any integrations and amounts to a change of gauge from
a  scalar potential to an effective vector potential description of the 
problem.

\end{document}